\documentclass[aps,prl,twocolumn,showpacs,floatfix,nobibnotes,superscriptaddress,amsmath,amssymb,longbibliography]{revtex4-1}

\usepackage{graphicx}
\usepackage{epsfig}
\usepackage{bm}
\usepackage{color}
\usepackage{float}
\usepackage{dcolumn}
\usepackage{multirow} 
\usepackage{hyperref}
\usepackage{amsmath}
\usepackage{amsfonts}
\usepackage{mathtools}
\graphicspath{{.}}

\newcommand{\be}{\begin{equation}}
\newcommand{\ee}{\end{equation}}
\newcommand{\ba}{\begin{eqnarray}}
\newcommand{\ea}{\end{eqnarray}}

\newcommand{\bra}[1]{\left\langle #1 \right|}
\newcommand{\ket}[1]{\left| #1 \right\rangle}
\newcommand{\expect}[1]{\langle {#1} \rangle}

\begin{document}

\title{Cloud Quantum Computing of an Atomic Nucleus} \thanks{This
  manuscript has been authored by UT-Battelle, LLC under Contract
  No. DE-AC05-00OR22725 with the U.S. Department of Energy. The United
  States Government retains and the publisher, by accepting the
  article for publication, acknowledges that the United States
  Government retains a non-exclusive, paid-up, irrevocable, world-wide
  license to publish or reproduce the published form of this
  manuscript, or allow others to do so, for United States Government
  purposes. The Department of Energy will provide public access to
  these results of federally sponsored research in accordance with the
  DOE Public Access
  Plan. (http://energy.gov/downloads/doe-public-access-plan).}

\author{E. F. Dumitrescu}
\affiliation{Computational Sciences and Engineering Division, Oak Ridge National Laboratory,
  Oak Ridge, TN 37831, USA}

\author{A. J. McCaskey}
\affiliation{Computer Science and Mathematics Division, Oak Ridge National Laboratory,
  Oak Ridge, TN 37831, USA}
  
\author{G.~Hagen}
\affiliation{Physics Division, Oak Ridge National Laboratory,
Oak Ridge, TN 37831, USA} 
\affiliation{Department of Physics and Astronomy, University of Tennessee,
Knoxville, TN 37996, USA} 

\author{G.~R.~Jansen}
\affiliation{National Center for Computational Sciences, Oak Ridge National
Laboratory, Oak Ridge, TN 37831, USA}
\affiliation{Physics Division, Oak Ridge National
Laboratory, Oak Ridge, TN 37831, USA}

\author{T.~D.~Morris}
\affiliation{Department of Physics and Astronomy, University of Tennessee,
Knoxville, TN 37996, USA} 
\affiliation{Physics Division, Oak Ridge National Laboratory,
Oak Ridge, TN 37831, USA} 

\author{T.~Papenbrock}
\email[Corresponding author: ]{tpapenbr@utk.edu}
\affiliation{Department of Physics and Astronomy, University of Tennessee,
Knoxville, TN 37996, USA} 
\affiliation{Physics Division, Oak Ridge National Laboratory,
Oak Ridge, TN 37831, USA} 

\author{R. C. Pooser}
\affiliation{Computational Sciences and Engineering Division, Oak Ridge National Laboratory,
Oak Ridge, TN 37831, USA}
\affiliation{Department of Physics and Astronomy, University of Tennessee,
Knoxville, TN 37996, USA} 

\author{D. J. Dean}
\affiliation{Physics Division, Oak Ridge National Laboratory,
Oak Ridge, TN 37831, USA}

\author{P. Lougovski}
\email[email: ]{lougovskip@ornl.gov}
\affiliation{Computational Sciences and Engineering Division, Oak Ridge National Laboratory,
Oak Ridge, TN 37831, USA}

\begin{abstract}
We report a quantum simulation of the deuteron binding energy on
quantum processors accessed via cloud servers. We use a Hamiltonian
from pionless effective field theory at leading order. We design a
low-depth version of the unitary coupled-cluster ansatz, use the
variational quantum eigensolver algorithm, and compute the binding
energy to within a few percent. Our work is the first step towards
scalable nuclear structure computations on a quantum processor via the
cloud, and it sheds light on how to map scientific computing
applications onto nascent quantum devices.
\end{abstract}

\maketitle {\it Introduction.---}Solving the quantum many-body problem
remains one of the key challenges in physics. For example,
wavefunction-based methods in nuclear
physics~\cite{caurier2005,navratil2009,barrett2013} face the
exponential growth of Hilbert space with increasing number of
nucleons, while quantum Monte Carlo
methods~\cite{koonin1997,lee2009,carlson2015} are confronted with the
fermion sign problem~\cite{Troyer:2005ui}.  Quantum computers promise
to reduce the computational complexity of simulating quantum many-body
systems from exponential to polynomial~\cite{nielsen2010}. For
instance, a quantum computer with about 100 error-corrected qubits
could potentially revolutionize nuclear shell-model
computations. However, present quantum devices are limited to about 20
non-error corrected qubits, and the implementation of quantum
many-body simulation algorithms on these devices faces gate and
measurement errors, and qubit decoherence. Nonetheless, the outlook
for quantum simulations is promising. A body of cutting edge research
is aimed at reducing computational complexity of quantum simulation
algorithms to match algorithmic requirements to the faulty
hardware~\cite{babbush2017}.

Recently, real-word problems in quantum chemistry and magnetism have
been solved via quantum computing using two to six
qubits~\cite{lanyon2010,peruzzo2014,omalley2016,kandala2017}. These
ground-breaking quantum computing experiments used phase estimation
algorithms~\cite{aspuruguzik2005} and the variational quantum
eigensolver (VQE)~\cite{peruzzo2014,mcclean2016}. They were performed
by a few teams of hardware developers working alongside
theorists. However, the field of quantum computing has now reached a
stage where a remote computation can be performed with minimal
knowledge of the hardware architecture. Furthermore, the relevant
software (e.g. {\tt PyQuil}~\cite{smith2016}, {\tt
  XACC}~\cite{maccaskey2017}, {\tt OpenQASM}~\cite{cross2017}, and
{\tt OpenFermion}~\cite{mcclean2017}) to run on quantum computers and
simulators is publicly available. Cloud access and cloud service to
several quantum processors now allows the broader scientific community
to explore the potential of quantum computing devices and algorithms.

In this Letter we present a quantum computation of the deuteron, the
bound state of a proton and a neutron, and we use only publicly
available software and cloud quantum hardware (IBM Q Experience and
Rigetti 19Q~\cite{otterbach2017}). The problem of quantum computing
the deuteron binding energy is still non-trivial because we have to
adjust the employed Hamiltonian, the wavefunction preparation, and the
computational approach to the existing realities of cloud quantum
computing. For example, the limited connectivity between qubits on a
quantum chip, the low depth (the number of sequential gates) of
quantum circuits due to decoherence, a limited number of measurements
via the cloud, and the intermittent cloud access in a scheduled
environment must all be taken into account.

This Letter is organized as follows. First, we introduce and tailor a
deuteron Hamiltonian from pionless effective field theory (EFT) such
that it can be simulated on a quantum chip. Next, we introduce a
variational wavefunction ansatz based on unitary coupled-cluster
theory (UCC)~\cite{mcclean2016,shen2017} and reduce the circuit depth,
and the number of two-qubit entangling operations, such that all
circuit operations can be performed within the device's decoherence
time. Next, we present the results of our cloud quantum computations,
performed on IBM QX5 and Rigetti 19Q quantum chips. Finally we give a
summary and an outlook.

{\it Hamiltonian and model space.---}Pionless EFT provides a
systematically improvable and model-independent approach to nuclear
interactions in a regime where the momentum scale $Q$ of the
interesting physics is much smaller than a high-momentum cutoff
$\Lambda$~\cite{vankolck1999,bedaque2002}. At leading order, this EFT
describes the deuteron via a short-ranged contact interaction in the
$^3S_1$ partial wave. We follow Refs.~\cite{binder2016,bansal2017} and
use a discrete variable representation in the harmonic oscillator
basis for the Hamiltonian. The deuteron Hamiltonian is
\be
\label{HN}
H_N = \sum_{n,n'=0}^{N-1} \langle n'|(T+V)|n\rangle a^\dagger_{n'} a_{n} .
\ee
Here, the operators $a^\dagger_{n}$ and $a_{n}$ create and annihilate
a deuteron in the harmonic-oscillator $s$-wave state $|n\rangle$.  The
matrix elements of the kinetic and potential energy are
\ba
\langle n'|T|n\rangle &=& {\hbar\omega\over 2}\bigg[ (2n+3/2)\delta^{n'}_n - \sqrt{n(n+1/2)}\delta_n^{n'+1}\nonumber\\
&&- \sqrt{(n+1)(n+3/2)}\delta_n^{n'-1}\bigg] , \nonumber\\
\langle n'|V|n\rangle &=& V_0 \delta_{n}^0\delta^{n'}_n .
\ea
Here, $V_0 = -5.68658111$~MeV, and $n,n'=0, 1, \ldots N-1$, for a
basis of dimension $N$.  We set $\hbar\omega=7$~MeV, and the potential
has an ultraviolet cutoff $\Lambda\approx 152$~MeV~\cite{konig2014},
which is still well separated from the bound-state momentum of about
$Q\approx 46$~MeV.

{\it Mapping the deuteron onto qubits.---}Quantum computers manipulate
qubits by operations based on Pauli matrices (denoted as $X_q$, $Y_q$,
and $Z_q$ on qubit $q$). The deuteron creation and annihilation
operators can be mapped onto Pauli matrices via the Jordan-Wigner
transformation
\begin{align}
a^\dagger_n  & \rightarrow  \frac{1}{2} \left[ \prod_{j=0}^{n-1}-Z_j \right] (X_n - i Y_n) , \nonumber\\ 
a_n & \rightarrow  \frac{1}{2} \left[ \prod_{j=0}^{n-1} -Z_j \right] (X_n + i Y_n) .
\end{align}
A spin up $\ket{\uparrow}$ (down $\ket{\downarrow}$) on qubit $n$
corresponds to zero (one) deuteron in the state $|n\rangle$. As we
deal with single-particle states, the symmetry under permutations
plays no role here. To compute the ground-state energy of the deuteron
we employ the following strategy. We determine the ground-state
energies of the Hamiltonian~(\ref{HN}) for $N=1,2,3$ and use those
values to extrapolate the energy to the infinite-dimensional space. We
have $H_1 = 0.218291 (Z_0 -I)$~MeV, and its ground-state energy
$E_1=\bra{\downarrow}H_1\ket{\downarrow}\approx -0.436$~MeV requires
no computation. Here, $I$ denotes the identity operation. For $N=2,3$
we have (all numbers are in units of MeV)
\begin{align}
H_2 &= 5.906709 I +0.218291Z_0 -6.125 Z_1 \nonumber\\
& -2.143304 \left(X_0 X_1 + Y_0Y_1\right) ,\label{H2}\\
H_3 &= H_2 + 9.625(I-Z_2) \nonumber\\
& -3.913119\left(X_1 X_2 +Y_1Y_2\right) \label{H3} .
\end{align}

For the extrapolation to the infinite space we employ the
harmonic-oscillator variant of L\"uscher's formula~\cite{luscher1985}
for finite-size corrections to the ground-state
energy~\cite{furnstahl2014}
\begin{eqnarray}
\label{extra}
E_N &=& -\frac{\hbar^2 k^2}{2m} \left(1 - 2{\gamma^2\over k} e^{-2k L} -4{\gamma^4 L\over k}e^{-4k L}\right)\nonumber\\
&& + {\hbar^2 k\gamma^2 \over m}\left(1-{\gamma^2\over k} -{\gamma^4\over 4k^2} +2w_2 k\gamma^4\right) e^{-4k L} .    
\end{eqnarray}
Here, the finite-basis result $E_N$ equals the infinite-basis energy
$E_\infty=-\hbar^2 k^2/(2m)$ plus exponentially small corrections. In
Eq.~(\ref{extra}), $L=L(N)$ is the effective hard-wall radius for the
finite basis of dimension $N$, $k$ is the bound-state momentum,
$\gamma$ the asymptotic normalization coefficient, and $w_2$ an
effective range parameter. For $N=1,2$ and $3$ we have $L(N) = 9.14$,
$11.45$, and $13.38$~fm as the effective hard-wall radius in the
oscillator basis with $\hbar\omega=7$~MeV, respectively, and
$L(N)\approx \sqrt{(4N+7)\hbar/(m\omega)}$ for $N\gg
1$~\cite{more2013}. Using the ground-state energies $E_N$ for $N=1,2$
allows one to fit the leading ${\cal O}(e^{-2kL})$ and subleading
${\cal O}(kLe^{-4kL})$ corrections by adjusting $k$ and
$\gamma$. Inclusion of the $N=3$ ground-state energy also allows one
to fit the smaller ${\cal O}(e^{-4kL})$ correction by adjusting
$w_2$. The results of this extrapolation are presented in the upper
part of Table~\ref{tab1}, together with the energies $E_N$ from matrix
diagonalization. We note that the most precise $N=2$ ($N=3$)
extrapolated result is about 2\% (0.5\%) away from the deuteron's
ground-state energy of $-2.22$~MeV.

\begin{table}[tb]
    \centering
    \begin{tabular}{|c|l|l|l|l|}\hline
    \multicolumn{1}{|c}{}  & \multicolumn{4}{c|}{$E$ from exact diagonalization}\\ \hline
    \multicolumn{1}{|c|}{$N$}  & \multicolumn{1}{c|}{$E_N$} & \multicolumn{1}{c|}{${\cal O}(e^{-2kL})$} & \multicolumn{1}{c|}{${\cal O}(kLe^{-4kL})$} & \multicolumn{1}{c|}{${\cal O}(e^{-4kL})$}\\ \hline
        2 & $-1.749$ & $-2.39$ & $-2.19$ & \\
        3 & $-2.046$ & $-2.33$ & $-2.20$ & $-2.21$ \\ \hline
    \multicolumn{1}{|c}{}  & \multicolumn{4}{c|}{$E$~from quantum computing}\\ \hline
    \multicolumn{1}{|c|}{$N$} & \multicolumn{1}{c|}{$E_N$}  & \multicolumn{1}{c|}{${\cal O}(e^{-2kL})$} & \multicolumn{1}{c|}{${\cal O}(kLe^{-4kL})$} & \multicolumn{1}{c|}{${\cal O}(e^{-4kL})$}\\ \hline
        2 &  $-1.74(3)$  & $-2.38(4)$ & $-2.18(3)$ &  \\
        3 &  $-2.08(3)$  & $-2.35(2)$ & $-2.21(3)$ & $-2.28(3)$ \\ \hline
        \end{tabular}
    \caption{Ground-state energies of the deuteron (in MeV) from
      finite-basis calculations ($E_N$) and extrapolations to infinite
      basis size at a given order of the extrapolation
      formula~(\ref{extra}). The upper part shows results from exact
      diagonalizations in Hilbert spaces with $N$ single-particle
      states, and the lower part the results from quantum computing on
      $N$ qubits. We have $E_1= -0.436$~MeV. The fit at ${\cal
        O}(e^{-4kL})$ requires three parameters and is only possible
      for $N=3$. The deuteron ground-state energy is $-2.22$~MeV.}
    \label{tab1}
\end{table}

\begin{figure}[tb]
    \centering
    \includegraphics[width=\columnwidth]{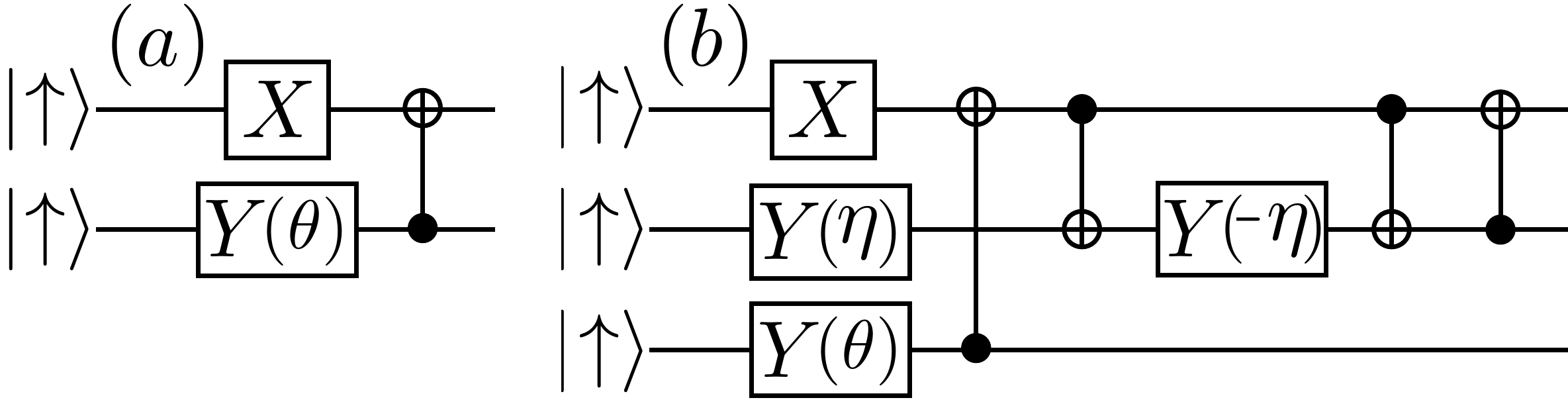} 
    \caption{Low-depth circuits that generate unitary rotations in
      Eq.~(\ref{u01}) (panel a) and Eq.~(\ref{u0102}) (panel b). Also
      shown are the single-qubit gates of the Pauli $X$ matrix, the
      rotation $Y(\theta)$ with angle $\theta$ around the $Y$ axis,
      and the two-qubit {\sc cnot} gates.}
    \label{circuits}
\end{figure}

{\it Variational wavefunction.---}In quantum computing, a popular
approach to determine the ground-state energy of a Hamiltonian is to
use UCC ansatz in tandem with the VQE
algorithm~\cite{mcclean2016,omalley2016,shen2017}. We adopt this
strategy for the Hamiltonians described by Eqs.~(\ref{H2}) and
(\ref{H3}). We define unitary operators entangling two and three
orbitals, \be
\label{u01}
U(\theta) \equiv e^{\theta\left(a_0^\dag a_1 - a_1^\dag a_0\right)} = e^{i{\theta\over 2}\left(X_0 Y_1 - X_1 Y_0\right)} , 
\ee
\ba
\label{u0102}
U(\eta, \theta) &\equiv& e^{\eta\left(a_0^\dag a_1 - a_1^\dag
  a_0\right) + \theta\left(a_0^\dag a_2 - a_2^\dag a_0\right)}
\\ \nonumber &\approx& e^{i{\eta\over 2}\left(X_0 Y_1 - X_1 Y_0
  \right)} e^{i{\theta\over 2}\left(X_0 Z_1 Y_2 - X_2 Z_1 Y_0\right)}.
\ea
In the second line of Eq.~(\ref{u0102}) we expressed the exponential
of the sum as the product of exponentials and note that the discarded
higher order commutators act trivially on the initial product state
$\ket{\downarrow \uparrow \uparrow}$. We seek an implementation of
these unitary operations in a low-depth quantum circuit.  We note that
$U(\eta)$ and $U(\eta, \theta)$ can be simplified further because a
single-qubit rotation about the $Y$ axis implements the same rotation
as Eq.~(\ref{u01}) within the two-dimensional subspace
$\{\ket{\downarrow \uparrow}, \ket{\uparrow \downarrow}\}$.  Likewise
Eq.~(\ref{u0102}) can be simplified by the above argument except the
first rotation now lies within the $\{ \ket{\downarrow \uparrow
  \uparrow}, \ket{\uparrow \downarrow \uparrow} \}$ subspace. The
second rotation, acting within the $\{ \ket{\downarrow \uparrow
  \uparrow}, \ket{\uparrow \uparrow \downarrow} \}$ subspace, must be
implemented as a $Y$-rotation controlled by the state of qubit 0 in
order to leave the $\ket{\uparrow \downarrow \uparrow}$ component
unmodified. The resulting gate decomposition for the UCC operations
are illustrated in Fig.~\ref{circuits}.

{\it Quantum computation.---}We use the VQE~\cite{peruzzo2014}
quantum-classical hybrid algorithm to minimize the Hamiltonian
expectation value for our wavefunction ansatz. In this approach, the
Hamiltonian expectation value is directly evaluated on a quantum
processor with respect to a variational wavefunction, i.e. the
expectation value of each Pauli term appearing in the Hamiltonian is
measured on the quantum chip. We recall that quantum-mechanical
measurements are stochastic even for an isolated system, and that
noise enters through undesired couplings with the environment. To
manage noise, we took the maximum of 8,192 (10,000) measurements that
were allowed in cloud access for each expectation value on the QX5
(19Q) quantum device. In contrast, the recent
experiment~\cite{kandala2017} by the IBM group employed up to $10^5$
measurements and estimated that $10^6$ would be necessary to reach
chemical accuracy on the six-qubit realization of the BeH$_2$ molecule
involving more than a hundred Pauli terms.  In addition to statistical
errors, we address systematic measurement errors by shifting and
re-scaling experimental expectation values as outlined in the
supplemental material of Ref.~\cite{kandala2017}. The expectation
values returned from the quantum device are then used on a classical
computer to find the optimal rotation angle(s) that minimize the
energy, or the parametric dependence of the energy on the variational
parameters is mapped for the determination of the
minimum~\cite{omalley2016}.

Our results are based on cloud access to the QX5 and the 19Q chips,
which consist of 16 and 19 superconducting qubits, respectively, with
a single qubit connected to up to three neighbors. This layout is well
suited for our task, because the Hamiltonian~(\ref{H3}) only requires
up to two connections for each qubit. We collected extensively more
data on the QX5 device than on the 19Q and only ran the $N=2$ problem
on the latter.

\begin{figure}[tb]
    \centering
    \includegraphics[width=\columnwidth]{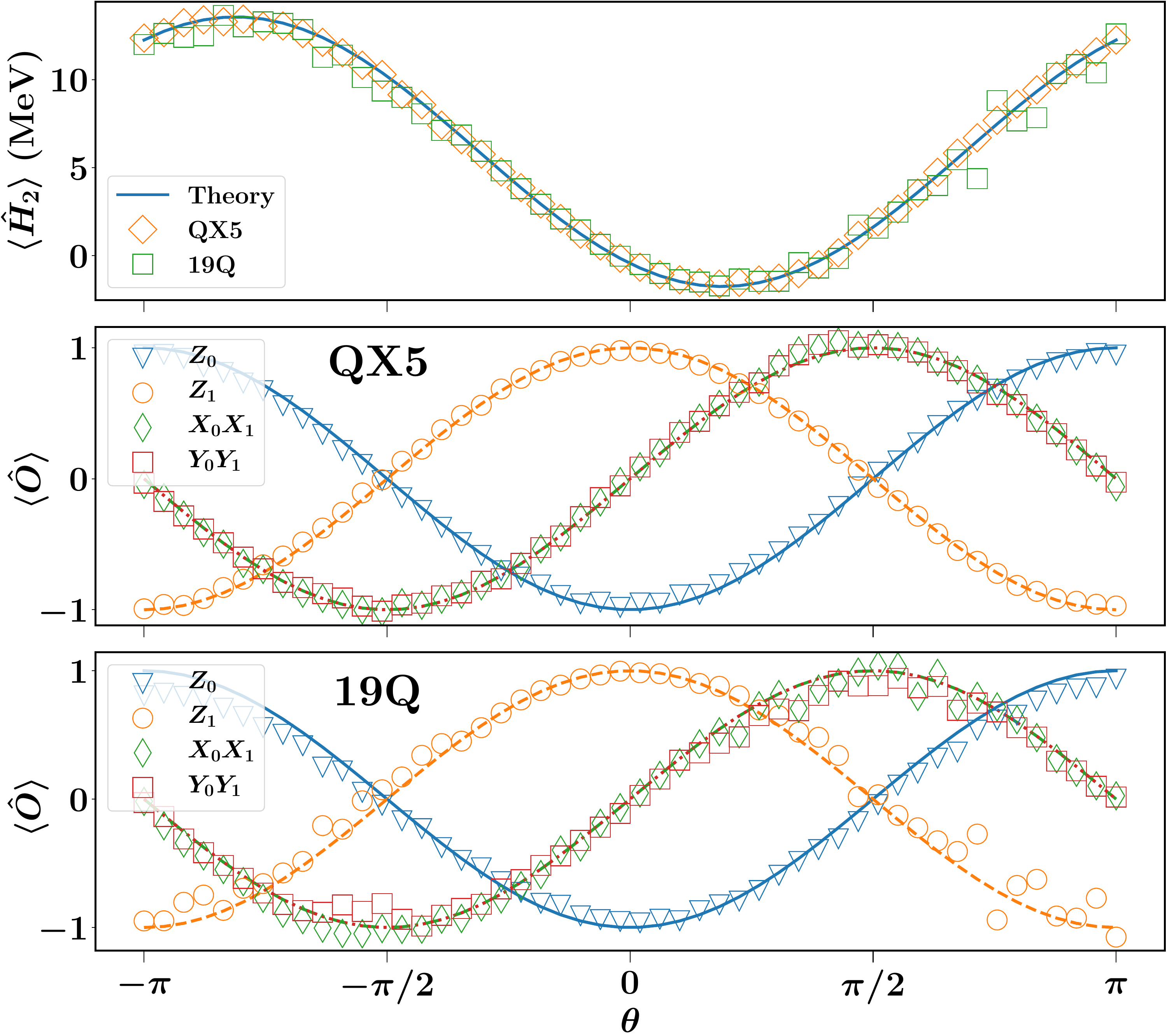} 
    \caption{(Color online) Experimentally determined energies for
      $H_2$ (top) and expectation values of the Pauli terms that enter
      the two-qubit Hamiltonian $H_2$ as determined on the QX5
      (center) and 19Q (bottom) chips. Experimental (theoretical)
      results are denoted by symbols (lines).}
    \label{2x2}
\end{figure}

{\it Results.---}Figure~\ref{2x2} shows $\langle H_2\rangle$ (top
panel) and the expectation values of the four Pauli terms that enter
the Hamiltonian $H_2$ as a function of the variational parameter
$\theta$ for the QX5 (center panel) and the 19Q (bottom panel). We see
that the measurements are close to the exact results, particularly in
the vicinity of the variational minimum of the energy. Cloud access,
and its occasional network interruptions, made the direct minimization
of the energy surface via VQE very challenging. Instead, we determined
the minimum energies $E_2^{\rm QX5}\approx -1.80\pm 0.05$~MeV and
$E_2^{\rm 19Q}\approx -1.72\pm 0.03$~MeV from fitting a cubic spline
close to the respective minimum.

Overall, the results obtained with the QX5 and 19Q quantum chips are
comparable in quality, keeping in mind the much larger access times we
had on the former device. Combining the independent results on both
chips yields $E_2= -1.74\pm 0.03$~MeV. This energy, as well as the
individual results, agree with the exact energy of $-1.749$~MeV within
uncertainties, see Table~\ref{tab1}.

To obtain the infinite-space result, we apply the leading and
subleading terms of the extrapolation formula Eq.~(\ref{extra}) to our
energies, i.e. $E_1=-0.436$~MeV and $E_2=-1.74\pm 0.03$~MeV, and
adjust $k$ and $\gamma$. The results for the extrapolated energy
$E_\infty=-{\hbar^2k^2/(2m)}$ are presented in Table~\ref{tab1} at
leading and subleading order of the extrapolation. They agree within
uncertainties with those from the exact diagonalization. Additionally,
the ${\cal O}(kL e^{-4kL})$ result of $-2.18(3)$~MeV deviates less
than 2\% from the exact deuteron ground-state energy of $-2.22$~MeV.

As a consistency check, we turn our attention to the $N=3$ case. These
quantum computations were only performed on the QX5. We optimized two
angles to find the minimum energy of the Hamiltonian in
Eq.~(\ref{H3}). We performed the minimization by choosing grids with
increasingly fine spacing in the parameter space around the minimum
(initially coarsely sampling the entire parameter space), computed the
energy expectation values on the quantum device, and determined the
minimum by a fit to cubic splines. This minimization problem is
significantly more challenging than for the $N=2$ case because the
increased number of {\sc cnot} gates introduced more noise and errors.

\begin{figure}[tb]
    \centering
    \includegraphics[width=\columnwidth]{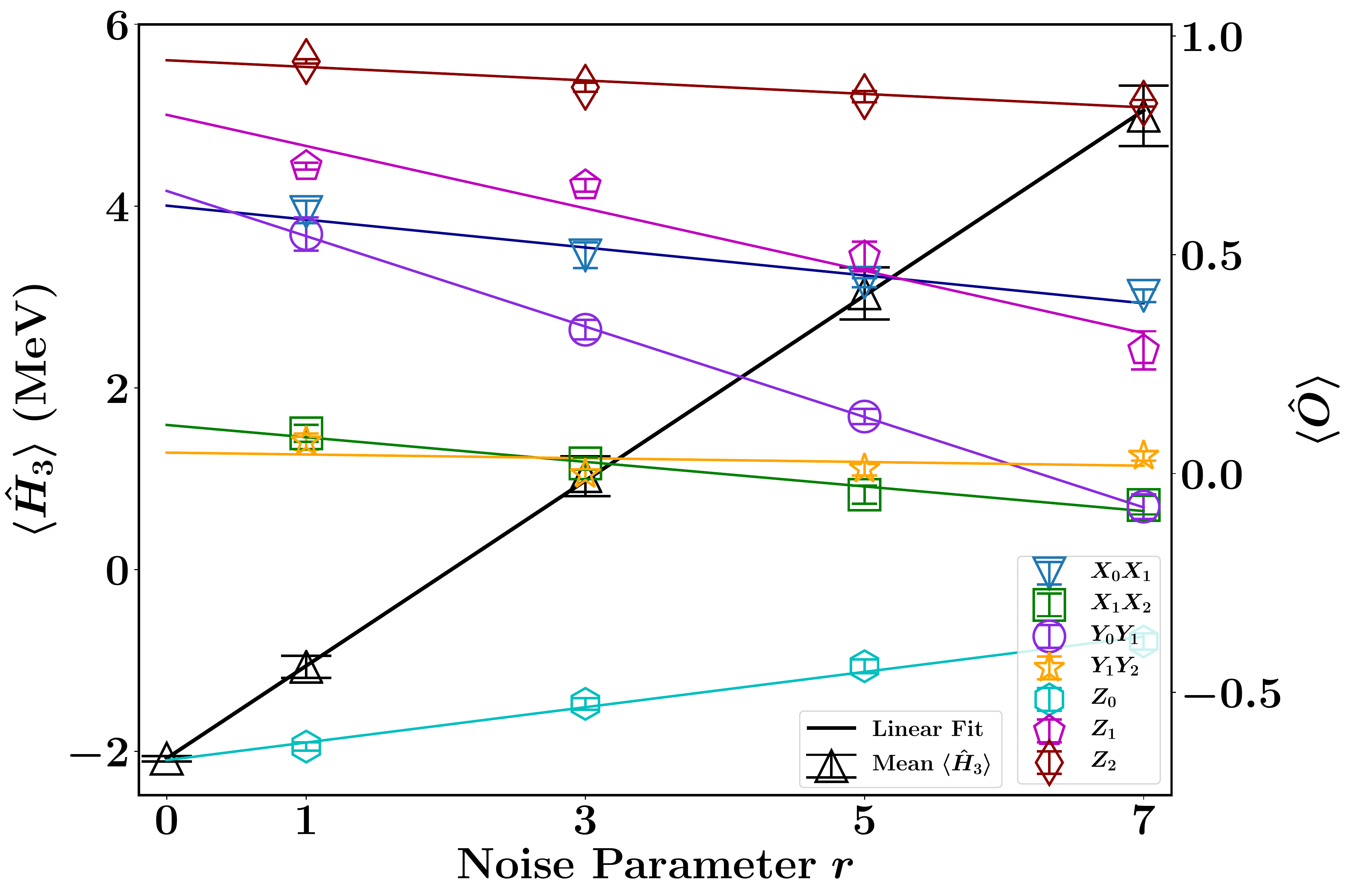} 
    \caption{(Color online) Noise extrapolation of the $N=3$ qubit
      problem run on the QX5. The $H_3$ energy (left axis, black line)
      and individual Pauli expectation values (right axis) are given
      as a function of the number {\sc cnot} gate scaling factor $r$.}
    \label{ibm3x3}
\end{figure}

In addition to correcting assignment errors, we implemented the
zero-noise extrapolation hybrid quantum-classical error mitigation
techniques~\cite{Li2017}.  For the zero-noise extrapolation, we
extrapolated the Hamiltonian expectation values $\expect{\hat{O}}$ to
their noiseless limit $\expect{\hat{O}}(0)$ with respect to noise
induced by the two-qubit {\sc cnot} operations. Since the true
entangler error model is not well established, we assume that a
generic two-qubit white noise error channel $\mathcal{E}(\rho) =
(1-\varepsilon)\rho + \varepsilon I/4$, where $\rho$ denotes the
density matrix, follows the application of each {\sc cnot}. We then
artificially increased the error rate $\varepsilon$ by adding pairs of
{\sc cnot} gates (i.e. noisy identity gates) to each {\sc cnot}
appearing in our original circuit. Our overall noise model is thus
parameterized by $r\varepsilon$, where $r$ is the number of {\sc cnot}
gate repetitions. A set of noisy expectation values
$\expect{\hat{O}}(r)$ are experimentally determined and used to
estimate their noiseless counterpart
$\expect{\hat{O}}(0)$~\cite{Li2017}. Kraus decomposing the white noise
channel in the two-qubit Pauli basis and noting that {\sc cnot} maps
the Pauli group onto itself, one can see that the noise channel
commutes and {\sc cnot} operators commute. After applying $r$ {\sc
  cnot}s (denoted by the operator $CX$), the noise channel becomes
$\mathcal{E}_{r}(\rho) = (1-r\varepsilon) CX \rho CX + r\varepsilon
I/4 + \mathcal{O}(\varepsilon^2)$. Given the small {\sc cnot} error
rate, quadratic contributions may be discarded and a linear regression
of the form $\expect{\hat{O}}(r) = \expect{\hat{O}}(0) + \chi r$, with
the slope $\chi = -\expect{\hat{O}}(0)\varepsilon$ yields noiseless
expectation values. Computing the individual Hamiltonian expectation
values for $r = 1,3,5,7$ using ten iterations of 8,192 measurements,
we then linearly extrapolated to the noiseless limit of $r=0$. This
approach yields a minimum energy $E_3=-2.08\pm 0.03$~MeV and agrees
with the exact result within uncertainties. Figure~\ref{ibm3x3} shows
the details of the noise extrapolations.

Finally, we include the $N=3$ results and apply the
extrapolation~(\ref{extra}) to find the infinite-space energy. The
results are shown in the lower part of Table~\ref{tab1}. We see that
the extrapolated energies agree with the exact results at the lowest
two orders. For the ${\cal O}(e^{-4kL})$ extrapolation, however, the
extrapolated energy yields about 3\% too much binding, and this
reflects the the differences between the $E_3$ value from the exact
and the quantum computation.

{\it Summary.---}We performed a quantum computation of the deuteron
binding energy via cloud access to two quantum devices. The
Hamiltonian was taken from pionless effective field theory at leading
order, and we employed a discrete variable representation to match its
structure to the connectivity of the available hardware. We adapted
the circuit depth of the state preparation to the constraints imposed
by the fidelity of the devices. The results from our two-qubit
computations on the IBM QX5 and the Rigetti 19Q devices agree with
each other and with the exact result within our small (a few percent)
uncertainties; the extrapolation to infinite Hilbert spaces yields a
result within 2\% percent of the deuteron's binding energy. Employing
a third qubit makes the computation more challenging due to
entanglement errors. Error correction methods again yield a deuteron
energy that agrees with exact results within uncertainties. The
extrapolation to infinite space is within 3\% of the exact result. The
presented results open the avenue for quantum computations of heavier
nuclei via cloud access.

\begin{acknowledgments}
  We acknowledge access to the IBM QX5 quantum chip, the Rigetti
  Quantum Virtual Machine, and to the Rigetti 19Q quantum chip. Plots
  were made with Matplotlib~\cite{Hunter2007}.  This material is based
  upon work supported by the U.S. Department of Energy, Office of
  Science, Office of Nuclear Physics, under grants DE-FG02-96ER40963,
  DE-SC0018223 (NUCLEI SciDAC-4 collaboration), and the field work
  proposals ERKBP57 and ERKBP72 at Oak Ridge National Laboratory
  (ORNL). This material is based upon work supported by the
  U.S. Department of Energy, Office of Science, Office of Advanced
  Scientific Computing Research (ASCR) quantum algorithms and testbed
  programs, under field work proposal numbers ERKJ332 and
  ERKJ335. This work used resources of the Oak Ridge Leadership
  Computing Facility located at ORNL, which is supported by the Office
  of Science of the Department of Energy under Contract No.
  DE-AC05-00OR22725.
\end{acknowledgments}

\end{document}